\documentclass[twocolumn]{aastex63}
\usepackage{color}

\newcommand       \msun        	{$M_{\odot}$}
\newcommand       \lsun      	{$L_{\odot}$}

\newcommand	     \myr              {$M_{\odot}$~yr$^{-1}$}

\newcommand        \mic        	 {$\mu$m}

\newcommand    \cii   {[C~{\footnotesize II}]}
\newcommand    \neii   {[Ne~{\footnotesize II}]}
\newcommand    \neiii  {[Ne~{\footnotesize III}]}
\begin{document}

\title{Dust masses, compositions, and luminosities in the nuclear disks and the \\ diffuse circumnuclear medium of Arp220}

\author{Eli Dwek}
\affiliation{Observational Cosmology Lab, NASA Goddard Space Flight Center, Mail Code 665, Greenbelt, MD 20771, USA}
\email{eli.dwek@nasa.gov}

\author{Richard G. Arendt}
\affiliation{CRESST~II/UMBC/GSFC}
\affiliation{Observational Cosmology Lab, NASA Goddard Space Flight Center, Mail Code 665, Greenbelt, MD 20771, USA}

\received{receipt date}
\revised{revision date}
\accepted{acceptance date}
\published{published date}
\submitjournal{The Astrophysical Journal}
\begin{abstract}
We present an analysis of the $4-2600$~\mic\ spectral energy distributions (SEDs) of the west and east nuclei and the diffuse infrared (IR) region of the merger-driven starburst Arp~220. We examine several possible source morphologies and dust temperature distributions using a mixture of silicate and carbonaceous grains.  From fits to the SEDs we derive dust masses, temperatures, luminosities, and dust inferred gas masses. We show that the west and east nuclei are powered by central sources deeply enshrouded behind $\sim 10^{26}$~cm$^{-2}$ column densities of hydrogen with an exponential density distribution, and that the west and east nuclei are optically thick out to wavelengths of $\sim 1900$ and $\sim 770$~\mic, respectively. The nature of the central sources cannot be determined from our analysis. We derive star formation rates or black hole masses needed to power the IR emission, and show that the \cii158\mic\ line cannot be used as a tracer of the star formation rate in heavily obscured systems. Dust inferred gas masses are larger than those inferred from CO observations, suggesting either larger dust-to-H mass ratios, or the presence of hidden atomic~H that cannot be inferred from CO observations. The luminosities per unit mass in the nuclei are $\sim 450$, in solar units, smaller that the Eddington limit of $\sim 1000 - 3000$ for an optically thick star forming region, suggesting that the observed gas outflows are primarily driven by stellar winds and supernova shock waves instead of radiation pressure on the dust.
\end{abstract}

\section{INTRODUCTION}

 Arp~220 is the ongoing merger between two galaxies \citep{joseph85,  armus87,sanders88}. Observations with the {\it Infrared Astronomical Satellite} ({\it IRAS}) showed it to be an ultra-luminous infrared galaxy (ULIRG) with a luminosity of about $(1-2)\times 10^{12}$~\lsun\  \citep{rangwala11}. Spectral observations with the {\it Infrared Space Observatory} ({\it ISO}) showed a lower \cii-to-IR luminosity ratio in ULIRGs, compared to normal star forming galaxies \citep{luhman03}. One of the explanations for this \cii\ deficit is that Arp~220 is optically-thick at far-infrared (FIR) wavelengths \citep[][and references therein]{fischer14}. Near-infrared (NIR) 2.2~\mic\ images revealed that the galaxy comprises two dusty nuclear systems embedded in a more diffuse dusty medium \citep[][{\it Hubble}/NICMOS]{graham90,soifer99}.  High spatial resolution with the Atacama Large Millimeter Array (ALMA) of the CO lines and continuum emission provided new insights into the gas dynamics, and the infrared (IR) luminosities and dust contents of the two nuclei, designated Arp220~West and Arp220~East \citep{rangwala11, wilson14, scoville17,wheeler20}. The ALMA continuum observations suggest that Arp~220~West is optically thick at 2600~\mic\ \citep{scoville17}. 

The energy sources powering the IR emission could either be bursts of intense star formation activity or active galactic nuclei (AGN). X-ray and radio continuum observations suggest that the IR luminosities are powered by intense star formation activities. Deep 0.5--8~keV band {\it Chandra} images of the inner 5\arcsec\ region and the detection of extended highly-ionized Fe~XXV line emission show the presence of outflows of hot thermal gas, presumably from merged supernova (SN) shocks and hot stellar winds \citep{paggi17}. Corroborative evidence is derived from high resolution observations of radio continuum emission, predominantly synchrotron emission attributed to expanding supernova remnants (SNRs) \citep{barcos-munoz15}. However the IR luminosity that can be attributed to star formation is constrained by the Eddington limit. Above a limiting luminosity surface density, radiation pressure on the dust will provide a feedback that reduces the  star formation activity of the region. Any luminosity above this limit must then be powered by an AGN \citep{scoville03,thompson05,iwasawa05,veilleux09, nardini10, contini13}.       
 In addition to the origin of the IR emission, the detailed properties of the two nuclei, their physical sizes, luminosities, gas and dust dust masses, as well as the inferred star formation rates (SFR) span a wide range of values, are yet to be determined.   
 
 In this paper we provide a detailed study of the dust masses and IR luminosities of the two nuclear systems and the ambient diffuse gas in Arp~220 in order to shed light on some of these open issues. Previous models of the IR continuum emission reproduced the peak emission of the galaxy, but did not model the separate contributions of the nuclei and the diffuse emission from the galaxy, or its silicate absorption feature \citep[][and references therein]{contini13}. Here we model the FIR SEDs of the two systems, as well as that of the diffuse circumnuclear medium over the entire $\sim 4 - 2600$~\mic\ wavelength region.  We also use physical dust properties to derive the dust masses and compositions.  
 Since there is strong evidence to suggest that the two nuclei are optically thick out to far-IR wavelengths, we  used simple models consisting of smooth distributions of dust and a predetermined temperature distributions to fit their observed SEDs. Our underlying assumption is that the dust temperature profiles in the nuclei are in a steady state that is determined by a balance between the absorption and reemission of the ambient radiation from any embedded stellar sources or a central BH. Models are characterized by different gas morphologies and dust temperature distributions and compositions. 

The paper is organized as follows. In Section~2 we summarize the relevant observational constraints on the nuclear and diffuse components of the galaxy. The model used to fit their SEDs and to derive their physical properties is described in Section~3. In Section~4 we present the model results, and a discussion of their astrophysical implications is given in Section 5.  

In our calculations we adopted a redshift of 0.018 to the galaxy \citep{ohyama99}, a Hubble constant of 70~km~s$^{-1}$~Mpc$^{-1}$, a value of $\Omega_m=0.30$, and a flat universe. The resulting luminosity and angular diameter distances to Arp~220 are 79 and 76~Mpc, respectively. An angular size of 0.1\arcsec\ then corresponds to a linear distance of 37~pc.
 
 \section{Observational constraints}
Observations of the continuum and CO lines emission have provided important information on the size, the dynamical mass, and the mass of the CO gas and the dust.    
The high-resolution ALMA observations of the kinematics of various CO, and HCN lines \citep{scoville17,wheeler20} provided important constraints on the dynamical masses and physical dimensions of the two nuclei.
Using an asymmetric disk model, \cite{wheeler20} derived nominal gas masses of $1.9\times10^9$, and $8\times 10^8$~\msun\ for Arp~220~West, and East, respectively. For a dust-to-H mass ratio $Z_{dH}=0.007$, the nominal value for the local diffuse ISM \citep{zubko04}, these gas masses translate to dust masses of $1.3\times 10^7$, and $5.6\times 10^6$~\msun, respectively. 

In their analysis, \cite{wheeler20} assumed that the density profiles of the disks fall off exponentially, with scale heights $a, b$ in radius and height. The values of $\{a,b\}$, in pc, derived in their analysis were $\{15.5, 35\}$ and $\{45, 20\}$ for, respectively, the west and east nuclei. An exponential sphere  containing the same mass as the disks will have an effective radius given by $R=0.8(a^2\, b)^{1/3} =16$ and 27~pc for the two nuclei, respectively. For comparison, the continuum 2600~\mic\ observations of \cite{scoville17} suggested that the approximately spherical distribution of the dust emission from the western nucleus extends out to a radius of 74~pc, whereas the emission from the eastern nucleus is more elongated, extending to a radius of about 111~pc.    
 
Using the nominal values of $\{M_d,R\}$ of  $\{1.3\times 10^7, 16\}$ and $\{5.6\times 10^6, 27\}$~$\{$\msun, pc$\}$, and a value of $\kappa(2600~\mu m)\approx 0.05$~cm$^2$~g$^{-1}$, we get radial optical depths of $\tau(2600\mu$m$)\approx 0.13$ and 0.021 for, respectively, the western and eastern nuclei. These values are significantly smaller than the optical depths of $\sim 1$ advocated in previous studies, meriting more detailed analysis of the far-IR to millimeter wavelengths SEDs of the two nuclei.   

\begin{figure}[t]
\includegraphics[width=3.0in]{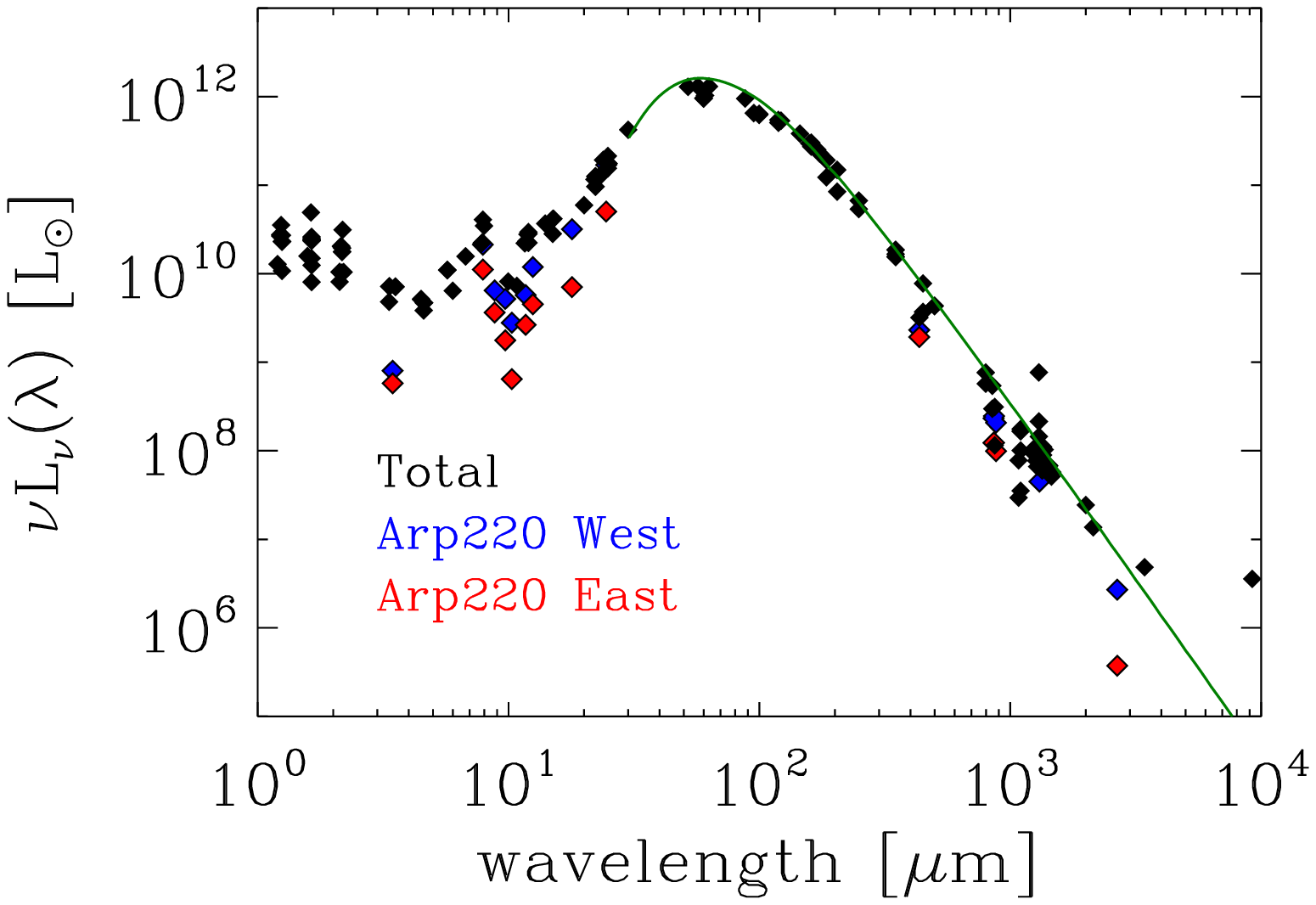}
\caption{\label{obs} Photometry for Arp 220 used to constrain models in this paper. Red and blue symbols are for the individual nuclei in observations with sufficient resolution to distinguish them.
The data, available in NED database, were compiled from the following  references: 
\citep{skrutskie06,sanders03,dunne00,dunne01b,veilleux09,soifer89,brown14,klaas01,spinoglio95,rangwala11,brauher08,forster04,moshir90,eales89,surace00,matsushita09,
chini86,martin11,rigopoulou96,sakamoto09,catalano14,imanishi07,lahius07,spinoglio02,anton04,carico92,sargsyan11,
gorjian04,thronson87,leroy11,truch08,zhang14,benford99,weedman09,brauher08}
}
\end{figure}

Figure~\ref{obs} presents the photometric observations of the different studies of Arp~220. The black diamonds represent the total flux from Arp~220, taken from the NED database. References are in the figure caption. The  blue and red diamonds represent the observations of Arp~220~West and Arp~220~East, respectively. The near to mid-IR 3.46 to 24.5 observations were taken from \citep{soifer99}, and the  430 to 2600~\mic\ observations were taken with the ALMA from \citep{scoville17}. We adopted their correction for the synchrotron contribution to the 2600~\mic\ emission. Fluxes at longer wavelengths are dominated by synchrotron radiation and were not considered in the fit to the thermal emission from the two nuclei.  

 \section{Model Description}
The primary input parameters that determine the IR emission from an optically thick cloud are its morphology, the nature and spatial distribution of the heating sources, and the dust composition. 

The spectrum from an IR-thick dusty source is represented by that of a blackbody, and  proportional to its area, so that all information on the dust mass giving rise to the emission is lost. However, most sources have an extended density profile. The optically thick region is then surrounded by an optically thin region, whose dust mass can be determined. If the optically thin region is part of a well defined density profile, then the mass of dust in the optically thick region can be inferred from knowledge of its mass in the outer thin region. This is the underlying principle allowing us to infer the dust mass in the optically thick nuclei of the galaxy.

\begin{figure}[t]
\includegraphics[width=3.0in]{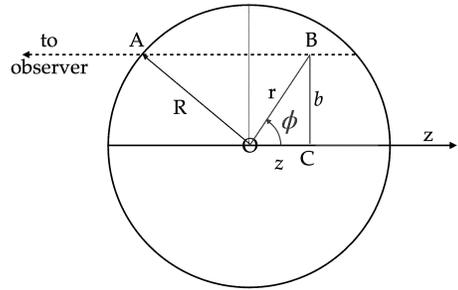}
\caption{\label{geo} A schematic figure showing the geometry and coordinate system used in the model.}
\end{figure}

\begin{figure*}[t]
\begin{center}
\includegraphics[width=3.0in]{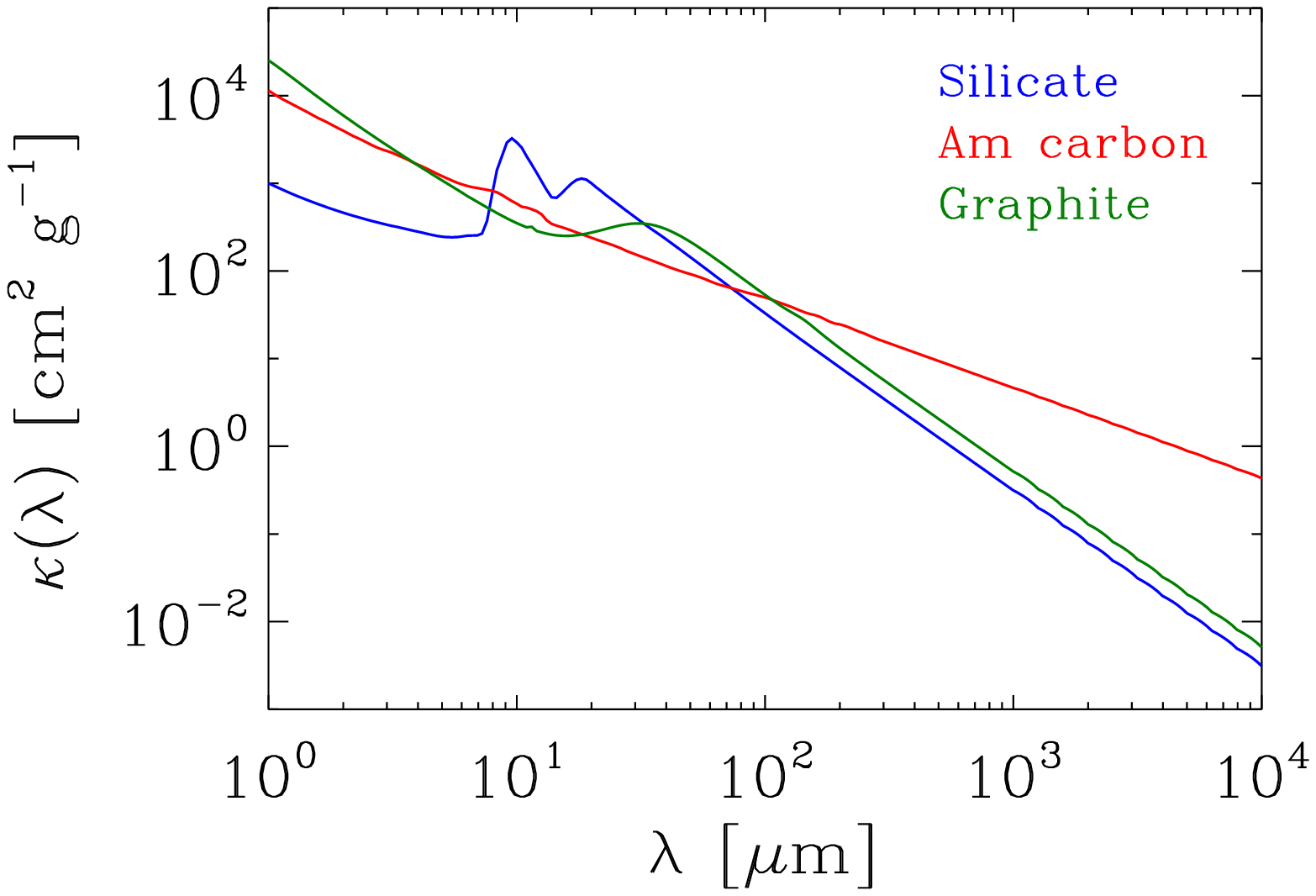}
\includegraphics[width=3.0in]{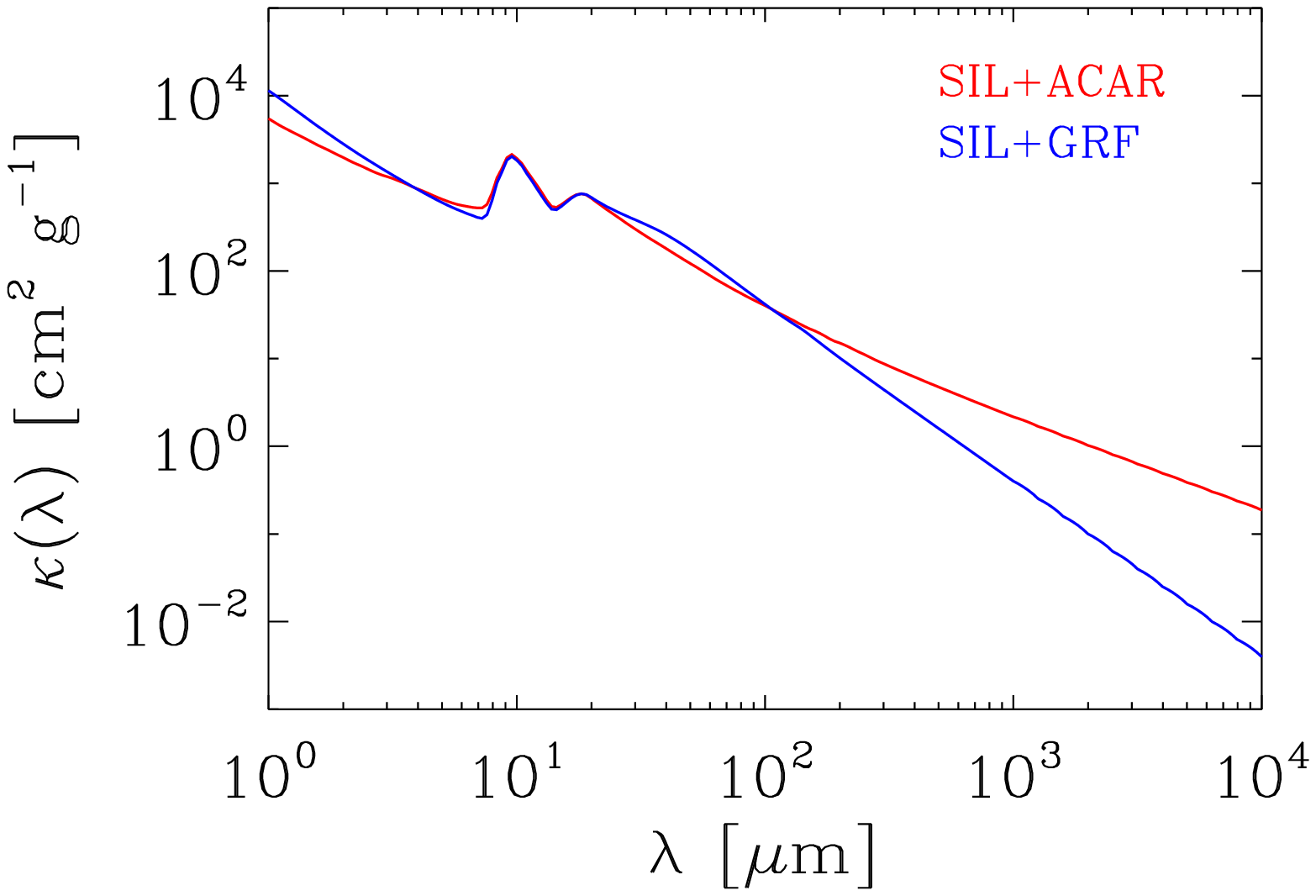}
\caption{\label{kappa} Left panel: The value of the dust mass absorption coefficient $\kappa(\lambda)$ for the three different dust compositions used in the paper. Right panel: The total mass absorption coefficient for the different compositional combinations of silicates and carbon dust.}
\end{center}
\end{figure*}

 We fit the SED  of the west and east nuclei with two emission components, a cold component to fit the emission at wavelengths longer than 20~\mic, and a warm component to fit the shorter wavelengths.

The cold component was taken to consist of a dusty sphere with either a uniform, an exponentially decreasing, or a power law dust density profile given, respectively, by 
\begin{eqnarray}
\rho_d(r)  & = & \rho_0 \\
 & = & \rho_0\, \exp(-r/a) \ \\
 & = & \rho_0\, (r/r_0)^{-\alpha} 
\end{eqnarray}
The characteristic density, $\rho_0$ in each case is determined by the total dust mass,
\begin{equation}
\label{md}
M_d = \int_0^R\ 4\pi\, \rho_d(r) r^2\, dr
\end{equation}
where $R$ is the radius of the sphere.
The dust mass column density, $\Sigma_d$ is given by
\begin{equation}
\label{sigd}
\Sigma_d = \int_0^R\ \rho_d(r)\, dr
\end{equation}

For each of these density profiles we considered three different temperature profiles. A uniform temperature profile, an exponentially decreasing one with the same scale length as the density profile, and a temperature profile characterized by an $r^{-1/2}$ power law. The first two profiles follow the spatial distribution of the dust in the appropriate cases, which is expected when the dust is heated by embedded stellar sources that are closely tied to the mass of the ambient dusty medium. The power law temperature distribution assumes that the radiation emerging from each radius within the sphere can be characterized by a blackbody at temperature $T_d$ and that, to first order, the total flux,  $\propto r^2\, T_d^4$, is conserved across each radius of the sphere. Such a temperature distribution is expected from dust that is heated by a central source. 

For sake of brevity we labeled each model by an abbreviated symbol DxTy, where either "x" or "y" take on the values of \{u, e, p\} for uniform, exponential or power-law, respectively, and the "D" and "T" refer to the quantity in question, density or temperature, respectively. So a model designated as DeTp is characterized by an exponential density profile and a power-law distribution of dust temperatures.
We fit the data with all possible density and temperature profile combinations, even though some combination are unphysical. For example, a uniform temperature profile requires the dust heating sources to be uniformly mixed throughout the region, which is not realistic for a source with an exponential or power law density profile. 

The unabsorbed emission from a volume element d$V$ of dust located at distance $r$ from the center of the sphere and radiating at temperature $T_d$ is given by
\begin{equation}
\label{Lnud}
dL_{\nu}(\lambda)= 4\, \rho_d(r)\, dV\, \pi\, B_{\nu}(\lambda, T_d)\, \kappa(\lambda) \qquad .
\end{equation}

For integration over the spherical distribution we choose a cylindrical
coordinate system $\{b,z,\theta\}$ (see Figure~\ref{geo}), where $\theta$ is the angle of rotation in the $xy$ plane (not shown in the figure). The observer
is located at a distance, $D\gg R$, along the negative $z$ axis. The volume
element in this system is given by $dV = db\, dz\, b\, d\theta$.

 The total emission from the cylinder is given by the integral
\begin{equation}
\label{lnu_cold}
dL_{\nu}(\lambda, b) = 8 \pi b\, db\, \int_{-z_b}^{z_b} \pi B_{\nu}[\lambda, T_d(r)]\, e^{-\tau(z,\lambda)} \kappa(\lambda)\, \rho_d(r)\,dz
\end{equation}  
where $z_b=|(R^2-b^2)^{1/2}|$, $r = (z^2 + b^2)^{1/2}$, and 
\begin{equation}
\label{tauz}
\tau(z, \lambda)= \kappa(\lambda)\, \int_{-z_b}^z\ \rho_d(r)\, dz
\end{equation}
To characterize the optical properties of the source, we define the optical depth along the z-axis between the observer and the layer at distance $r$ from the center, $\tau(r,\lambda)$, as 
 \begin{equation}
\label{taurad}
\tau_(r,\lambda)= \kappa(\lambda)\, \int_{-R}^r\ \rho_d(z)\, dz \qquad ,
\end{equation}
and the radial optical depth, $\tau_R(\lambda)$, of the sphere along the z-axis as 
\begin{equation}
\label{taur}
\tau_R(\lambda) \equiv \tau(0,\lambda) = \kappa(\lambda)\, \int_{-R}^0\ \rho_d(z)\, dz \qquad ,
\end{equation}
 
Initial calculations showed that the short wavelength emission and strong 10~\mic\ absorption feature could not be reproduced by the smooth dust emission component giving rise to the cold emission. We therefore assumed the existence of a separate component giving rise to the short wavelength emission.
The lack of any significant UV-optical emission from the galaxy suggested that any stellar or AGN emission is absorbed by a UV-optically thick dust and reradiated at IR wavelengths. We assumed that the absorbing dust is optically thick at IR wavelengths as well. The warm component was therefore taken to consist of a central source with a blackbody of radius $R_{bb}$ and a dust temperature $T_{bb}$, and intervening cold dust characterized by an optical depth $\tau_{cold}(\lambda)=\kappa(\lambda)\, \Sigma_d$, where $\Sigma_d$ is the mass surface density of the intervening dust. We assumed that any reradiated emission from the intervening dust makes a negligible contribution to the emission from the galaxy.
The emission from the warm component is then given by
\begin{equation}
\label{lnu_warm}
L_{\nu}(\lambda) = 4 \pi \, R_{bb}^2 \, \pi B_{\nu}(\lambda, T_{bb})\, e^{-\tau_{cold}(\lambda)} 
\end{equation}  
The model of each nucleus is thus characterized by 6 parameters: $M_d, T_d, X, R_{bb}, T_{bb}, \Sigma_d$, where the parameter $X$ is equal to $R$, the radius of the sphere with the uniform or the power law density profile, or equal to $a$, the scale length of the sphere with the exponential density profile. 
The diffuse emission component was assumed to arise from an optically thin distribution of dust with mass $M_d$, radiating, at a single dust temperature $T_d$:
\begin{equation}
\label{lnu_diff}
L_{\nu}(\lambda)=4\, M_d\, \pi\, B_{\nu}(\lambda, T_d)\, \kappa(\lambda) \qquad .
\end{equation}

We considered the dust composition in the nuclei and the diffuse medium to consist of a mixture of either silicate and amorphous carbon (SIL+ACAR) or a mixture of silicate and graphite (SIL+GRF) grains. 
We adopted a silicate-to-carbon dust mass ratio of 4:3, representing that of the dust in the local interstellar medium (ISM) of the Milky~Way \citep{zubko04}.  Figure~\ref{kappa} (left panel) shows the value of the dust mass absorption coefficient for each of the dust compositions considered in the paper. The right panel of the figure shows the mass absorption coefficient for the two adopted dust mixtures. The value of $\kappa$ is largely independent of the grain size, since at all wavelengths and grain radii of interest, ($\lambda \gtrsim 1~$\mic,\ $  a_{gr} \lesssim 1$~\mic) the size parameter, $x=2\pi a_{gr}/\lambda$, is in the Rayleigh region in the Mie calculations.  We also constrained the relative temperatures of the silicate and carbon dust. Exposed to, or immersed in, the same radiation field, the silicate and carbon (ACAR or GRF) grains attain different temperatures with an  approximate silicate-to-carbon temperature ratio of 1:1.25 \citep{zubko04}. All dust temperatures quoted in this paper correspond to that of the silicate dust component.

\section{Model results}
 
 Model fits to the data show that the component giving rise to the long wavelength ($\lambda > 20$ $\mu$m) 
emission for both nuclei was best represented by a sphere with an exponential density profile. 
However, with this density profile, several different temperature distributions gave similarly good fits to the observed SED.
To discriminate between these temperature profiles we also calculated the expected (beam-convolved) brightness profiles 
for each model at 2600 $\mu$m and compared to the observed intensity profiles presented by (Scoville et al. 2017, Figure 1). 
The results are shown in Figure 4. The flux conserving $T_{\rm d}(r) \sim r^{-0.5}$ profile (model DeTp) provided the 
best fit to the west nucleus. 
However, the east Nucleus was best represented by the (non-physical) uniform temperature distribution (model DeTu).
Despite this, we adopted model DeTp as the representative model for the east nucleus as well, since observed elongation
of the east nucleus may account for discrepancies between the observed brightness profile and the one predicted 
by a spherical model. 
Both the DeTp and DeTu models provide similarly good fits to the SED of the east nucleus. 

\begin{figure*}[t]
\includegraphics[width=3.0in]{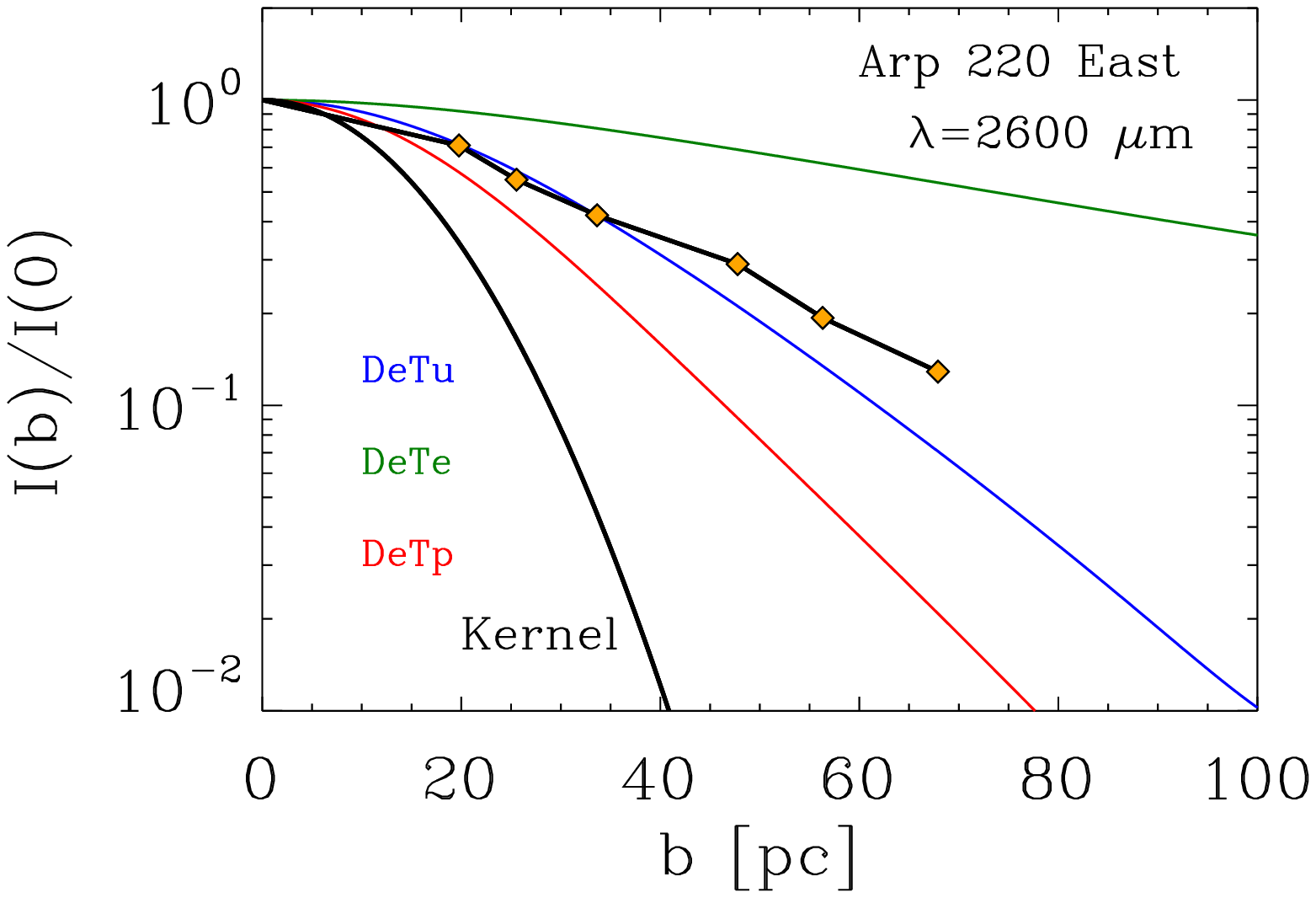}
\includegraphics[width=3.0in]{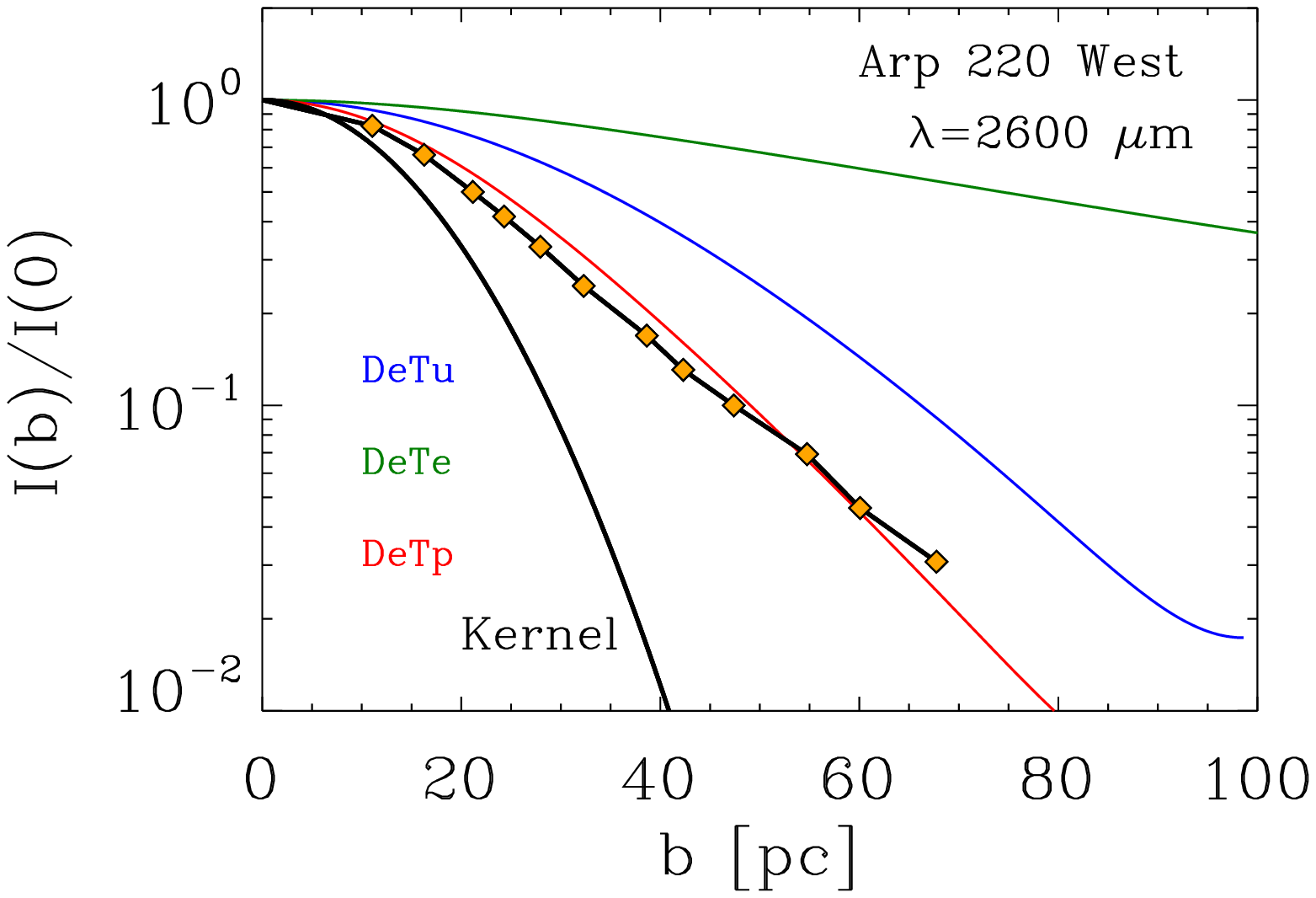}
\caption{\label{Fconvol} Comparison of the beam-convolved 2600~\mic\ intensity profiles  (solid lines) to observation (diamond symbols). The parameter $b$ is the impact parameter (Fig. \ref{geo}). The solid black line is the profile of the convolution kernel or beam.} 
\end{figure*}
 
\begin{figure*}[t]
\begin{center}
\includegraphics[width=2.2in]{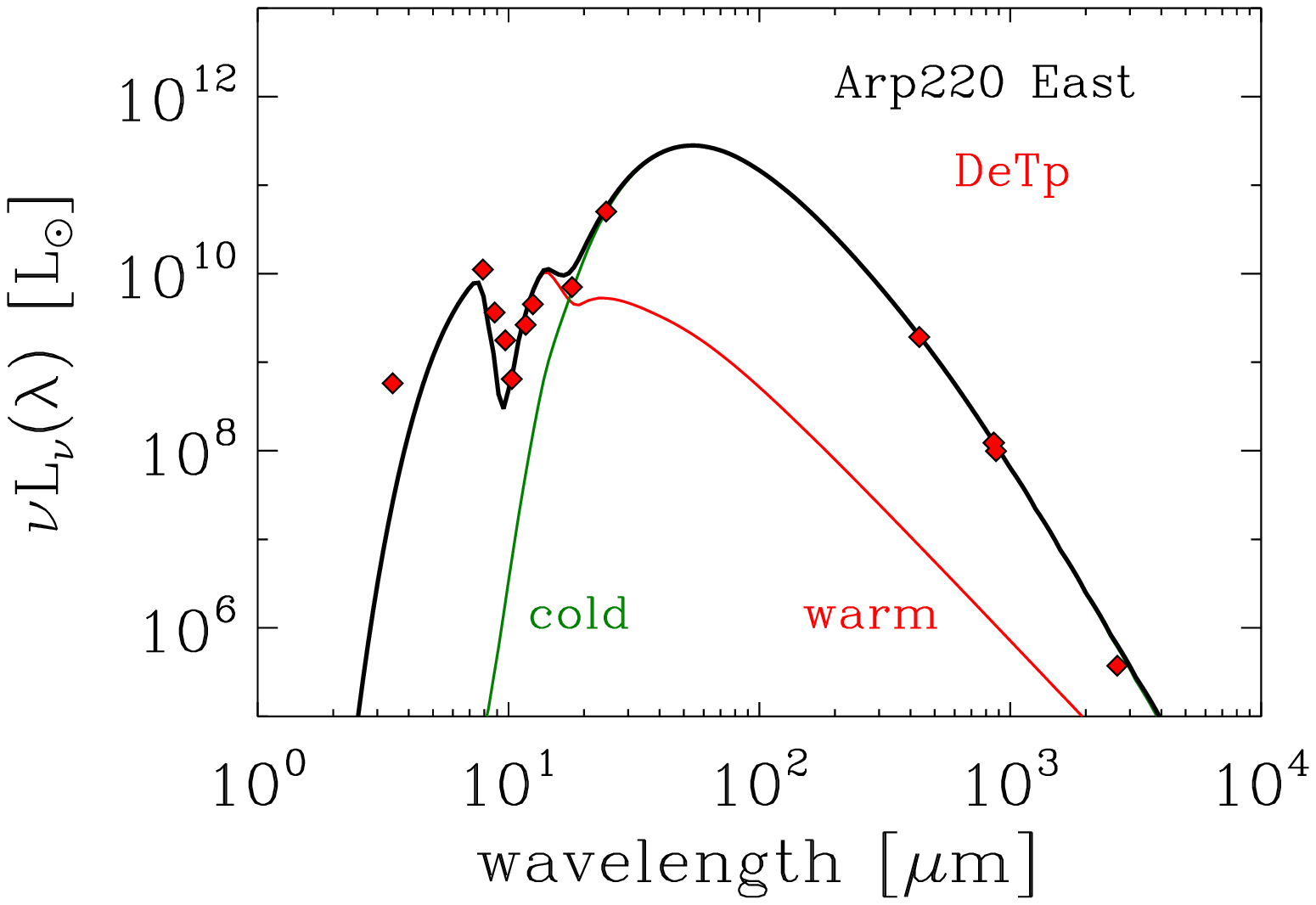}
\includegraphics[width=2.2in]{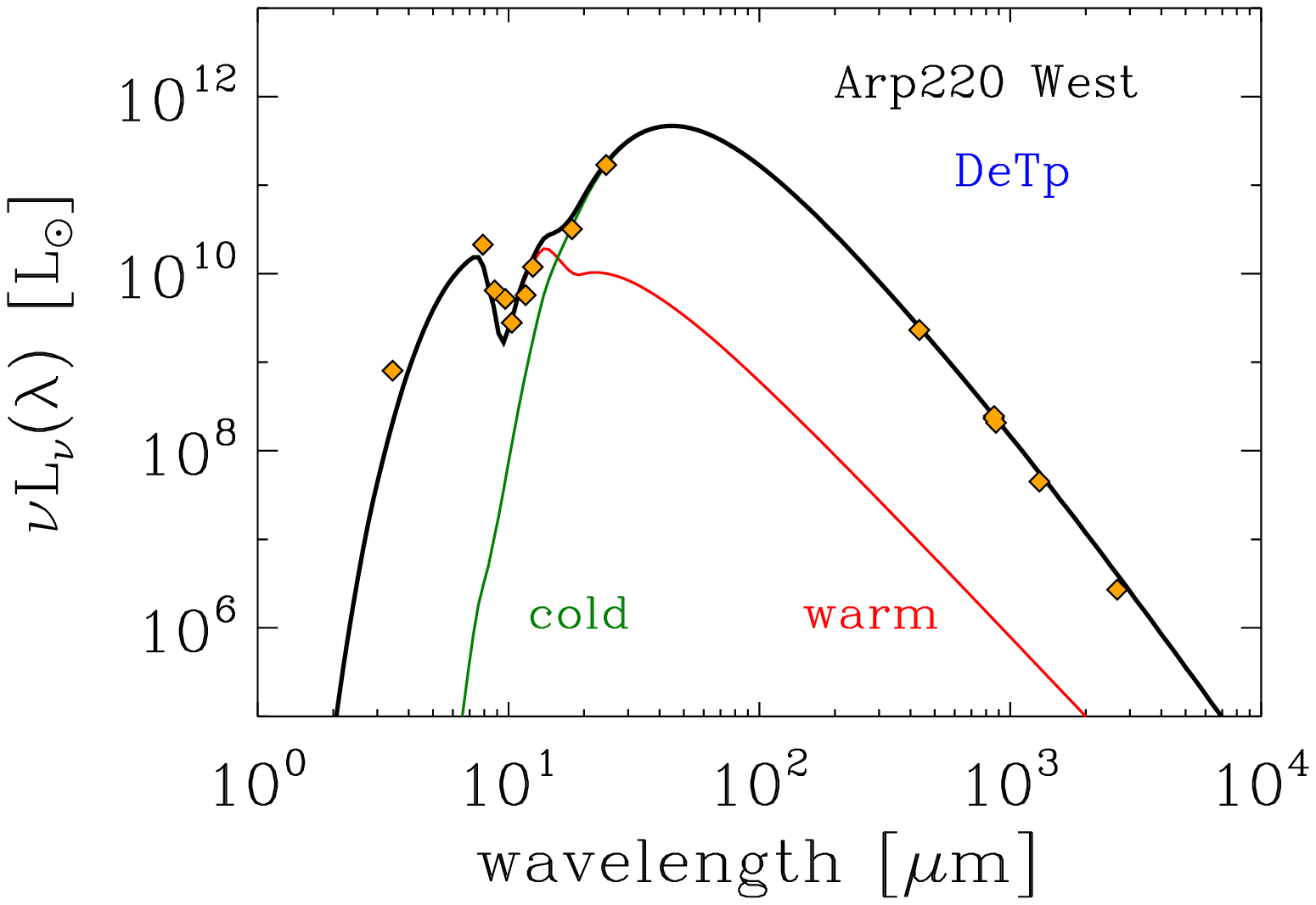}\\ 
\includegraphics[width=2.2in]{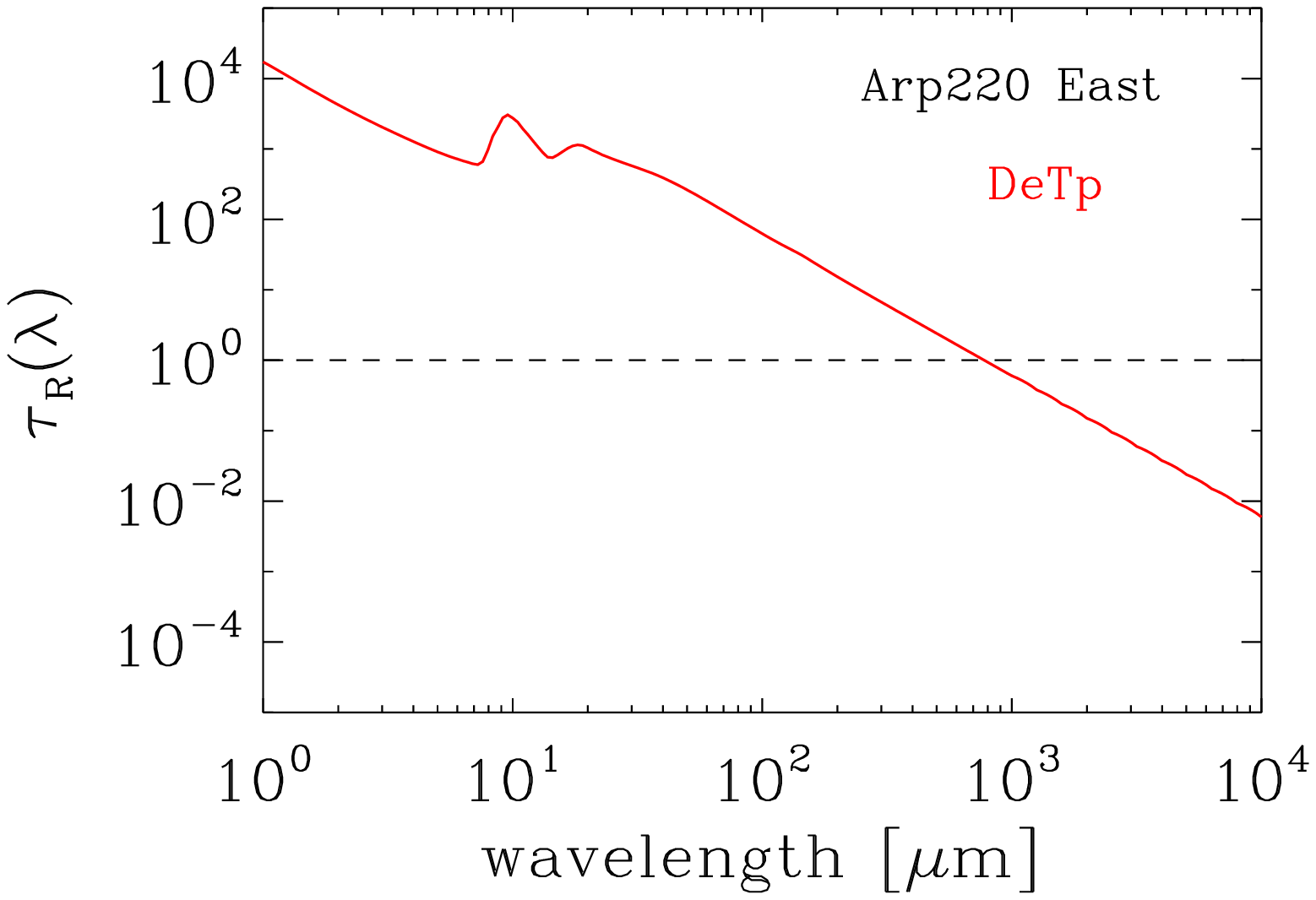}
\includegraphics[width=2.2in]{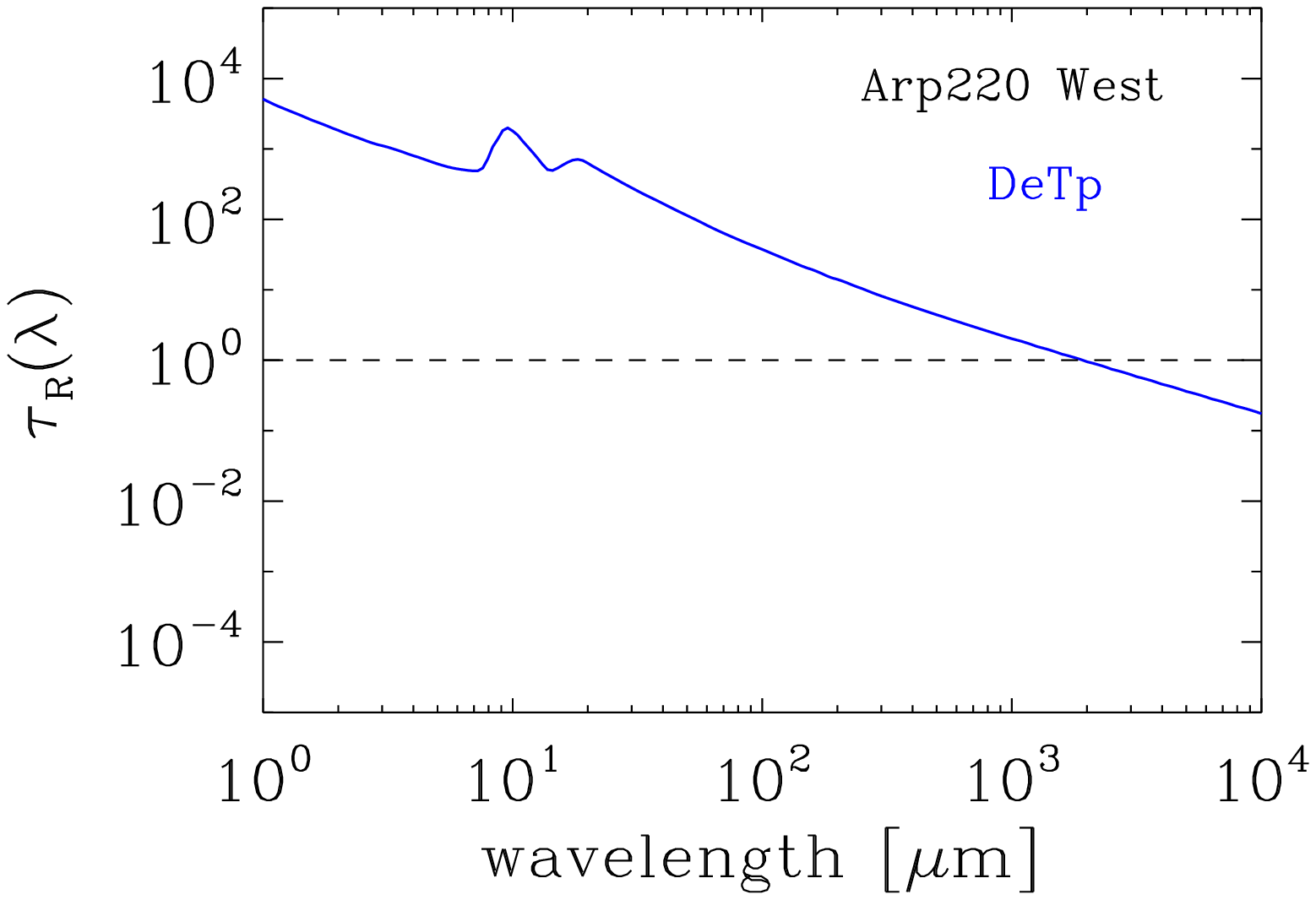}\\
\includegraphics[width=2.2in]{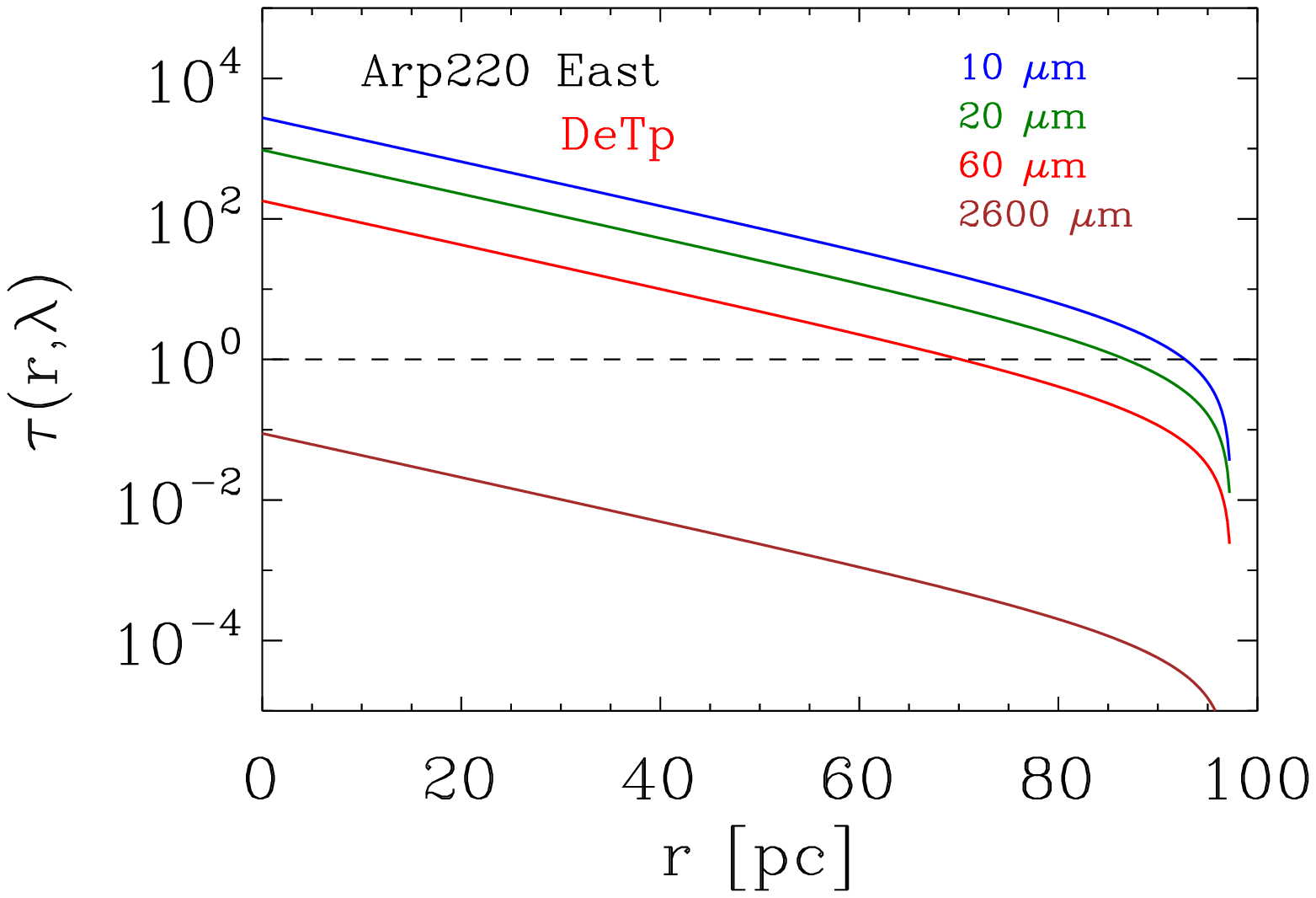}
\includegraphics[width=2.2in]{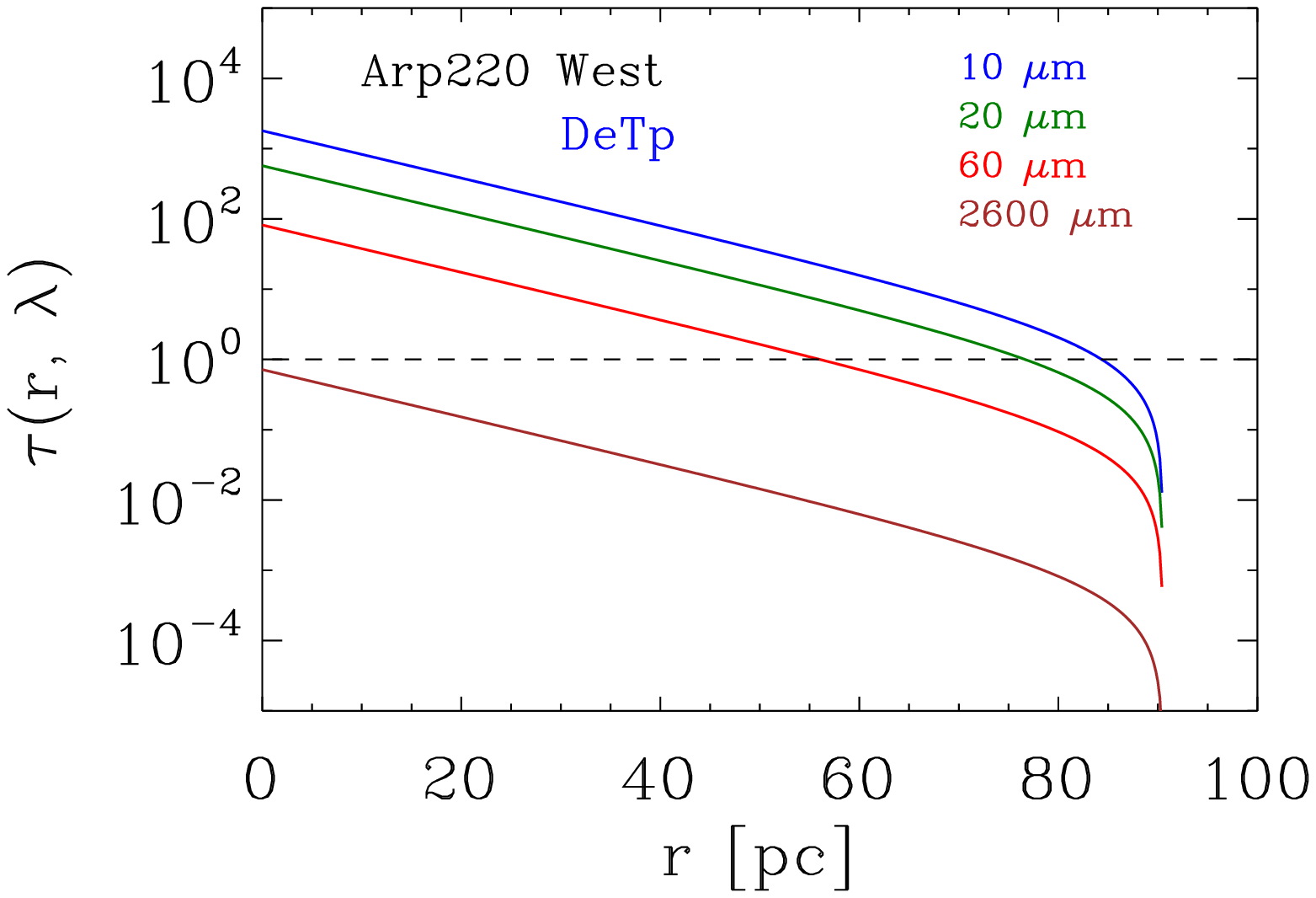} \\
\caption{\label{sedfits}  Results of the best fit models to the data for Arp~220~East and West. Shown are the spectral fits (top row), the wavelength dependence of the radial opacity $\tau_R(\lambda)$ (middle row), and radial dependence of the intervening opacity $\tau(r,\lambda)$ from $r$ to the observer (bottom row). The fits required a dust composition consisting of SIL+ACAR dust for the west, and one consisting of SIL+GRF dust for the east nucleus.}
\end{center}
\end{figure*}

Figure~\ref{sedfits} presents the best fitting physical model to the data for each nucleus. 
 The results for the east and west nuclei are presented in the left and right column, respectively. 
The top row shows the fit to the spectra with the contribution to the cold and warm components represented by green and red curves, respectively. 

The middle row shows the dependence of the radial opacity $\tau_R(\lambda)$, defined in eq. (\ref{taur}) as a function of wavelength. The figure shows that the west and east nuclei are optically thick at wavelengths $\lambda \leq 1890$ and $\lambda \leq 775$~\mic, respectively.
It is customary to compare the optical depth of some astronomical sources, e.g. supernovae interiors \citep{clayton74} or the Arp~220 nuclei \citep{scoville17}, to that of a slab of concrete. We find that the radial optical depth to the nuclei at these wavelengths is equivalent to that of $\sim 1$~cm thick slab of concrete with a dust mass column density of $\sim 3$~g~cm$^{-2}$ at millimeter wavelengths. 

The bottom row depicts $\tau(r, \lambda)$, defined as the intervening radial optical depth between the dust layer at radius $r$ and the observer (see eq.~(\ref{taurad})). The figure illustrates the depth to which the source can be probed at different wavelengths. It shows that the sources are mostly transparent at 2600~\mic. 
The west nucleus can be probed down to radii of 84, 76, and 56~pc at 10, 20, and 60~\mic, respectively, until a surface of unit optical depth is encountered, and the east nucleus can be probed down to radii of 92, 87, and 70~pc at these wavelengths, respectively. The radial optical depths at 2600~\mic\ are 0.09 and 0.72 for the east and west nuclei, respectively. These values are consistent with those derived by \citep{scoville17}. 
The parameters of the fit, $M_d, T_d, a$  of the west and east nuclei are listed in Table~1.  
 
 The parameters of the fit to the warm emission component, $R_{bb}, T_{bb}$ and the mass column density of the intervening cold dust, $\Sigma_d$ are also listed in Table~1.  The mass column density required to create the silicate absorption feature is about $10^3$ lower than that attenuating the nuclear emission. The warm component could therefore be a single intervening clump of dust along a relatively transparent line of sight to the source. Since the geometry of the intervening dust is unknown, one cannot derive its mass. The total luminosity absorbed by the intervening dust is given by $L_{bb}-L_d \approx 4\times 10^{10}$~\lsun, where $L_d$ is the observed IR luminosity emitted at wavelengths $\lesssim 40$~\mic. We assumed that the absorbed emission makes a negligible contribution to the total SED of the galaxy.

\section{Astrophysical implications}
Table~1 lists several derived properties from the spectral fits that have important implications for the morphology of the Arp~220 nuclei, the nature of their heating sources, and the composition and survival of dust in galaxy mergers.

\subsection{Dust masses and Dust-inferred gas masses}
Dust masses in our models were derived using the optical properties of known astrophysical constituents. The total dust mass in the galaxy is about $1.3\times 10^8$~\msun, similar to the estimate of $\sim 10^8$~\msun given by \cite{rangwala11}. 
The derived dust masses can be used to estimate the associated mass of molecular plus atomic gas. For a dust-to-H mass ratio of 0.007, the inferred gas masses are $2.6\times10^9$ and $4.9\times10^9$~\msun\ for the west and east nuclei, respectively, and $1.1\times10^{10}$ for the diffuse component of the galaxy. 
 
These dust derived gas masses are significantly higher than the mass of the molecular gas derived from CO observations. Using a CO to H$_2$ conversion factor of $X=0.8$~\msun (K~km~s$^{-1}$~pc$^2$)$^{-1}$, \cite{wheeler20} derived H$_2$ masses of 
 $(0.8-2.4)\times10^9$ and $(0.4-0.8)\times10^9$~\msun\ for the west and east nuclei respectively. These values are similar to the values of $>1.4\times10^9$ and $>0.4\times10^9$ derived by \cite{scoville17} who may, however, used a higher $X$-factor for the CO to H$_2$ mass conversion.
The higher dust-inferred gas masses suggest that a fraction of the gas in the nuclei may be in atomic form, much more so in Arp~220~East. Alternatively, if we take the inferred H$_2$ mass as the representative gas mass, then  the dust-to-H mass ratio, $Z_{dH}$, for the west and east nuclei is 0.011 and 0.082, respectively, significantly higher than the value 0.007 in the local ISM.  
 
\subsection{Luminosities}
\begin{figure*}[t]
\begin{center}
\includegraphics[width=3.0in]{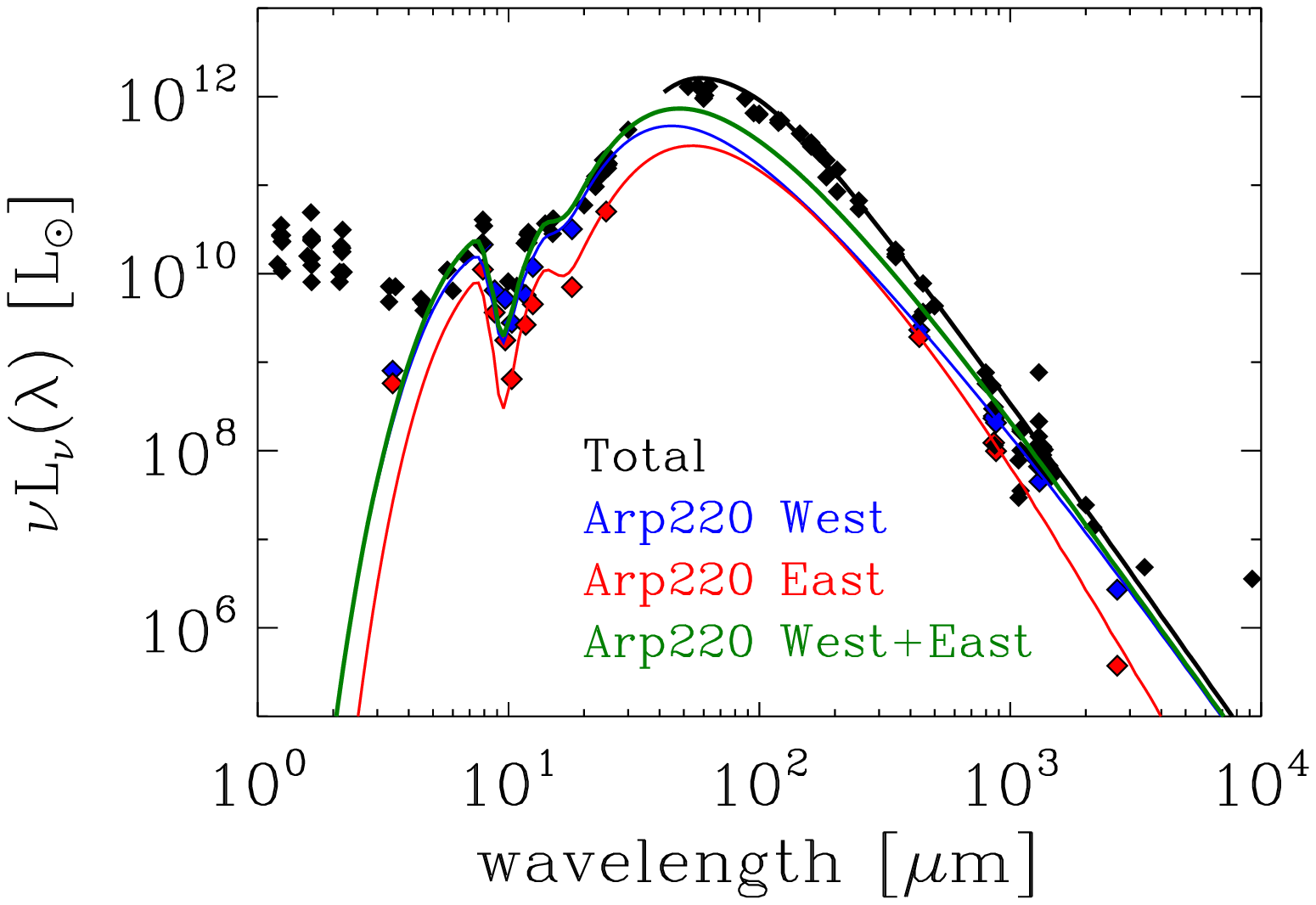}
\includegraphics[width=3.0in]{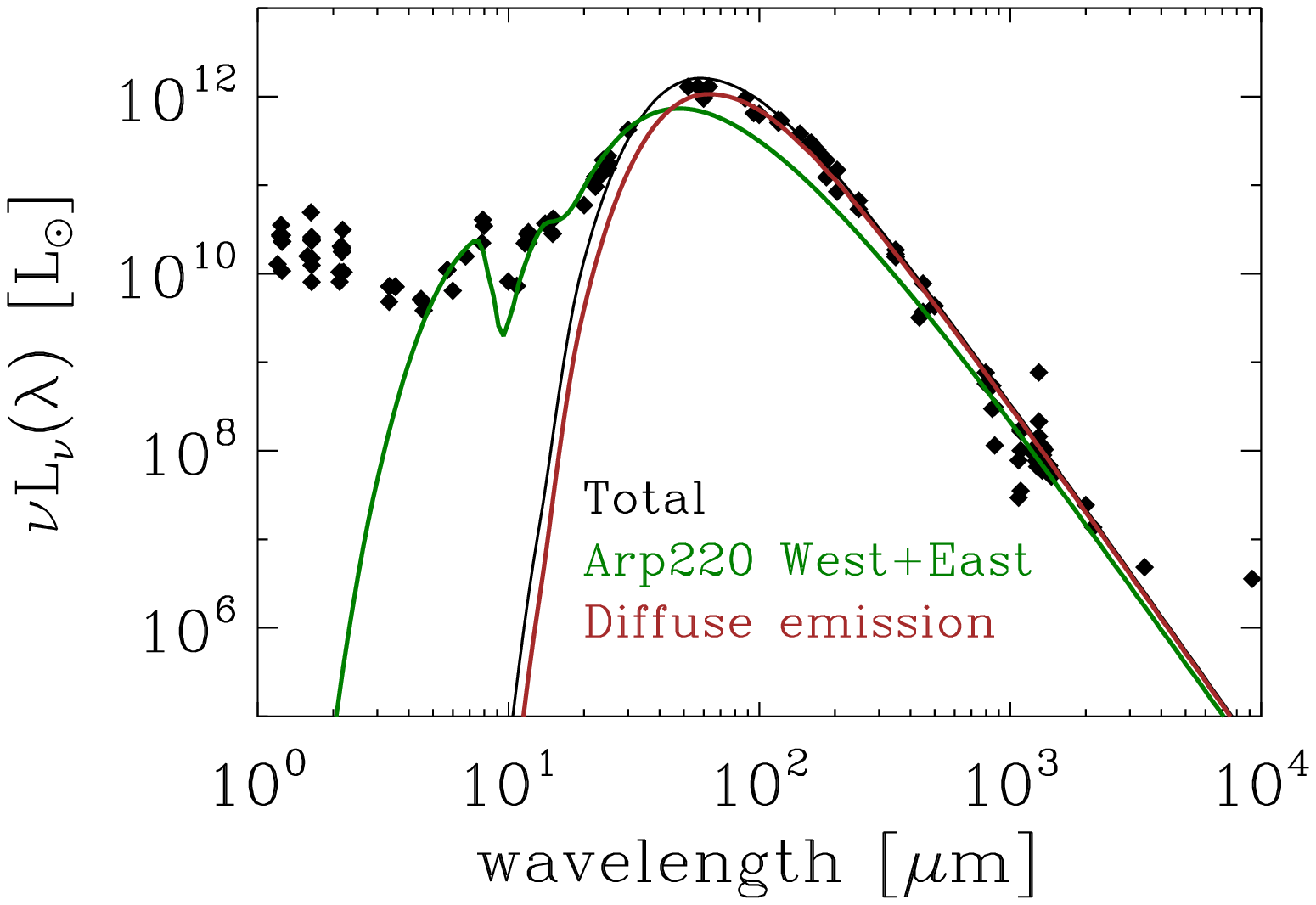}
\caption{\label{spectot} Left panel:  A comparison of the model spectra to the observations. The blue and red curves represent the sum of the warm and cold components to the emission from the west and east nuclei, respectively. The green curve is the total emission from the nuclear region. The blue, red and black diamonds are as in Figure~\ref{obs}. The black line is the fit to the total SED at wavelengths $> 40$~\mic. Right panel: Same total photometry as the left panel with the combined nuclear and diffuse fluxes, shown in green and brown colors, respectively.}
\end{center}
\end{figure*}

The morphology and temperature distribution of the nuclei suggests that they are powered by a central source, which could be either an accreting BH or an active region of star formation. 
Figure~\ref{spectot} compares the fluxes from the nuclear regions of the galaxy to the total observed SED. 
The diffuse emission component is obtained by fitting the total SED after subtraction of the contributions from the east and west nuclei. At wavelengths longer than 30~\mic\ it can be fitted with a SIL+ACAR mixture of dust radiating at a temperature of 40 K and a luminosity and dust mass of about $9.5\times10^{11}$~\lsun\ and $8\times10^7$~\msun, respectively. The total luminosity of Arp~220 is $1.9\times10^{12}$~\lsun, similar to the value of $(1-2)\times 10^{12}$~\lsun\ derived in previous studies \citep[e.g.][]{sanders03, armus09, wilson14, scoville17}. The luminosity of the west nucleus is consistent with the lower limit of $3\times10^{11}$~\lsun\ obtained by \cite{sakamoto08}, and almost identical to the value of $6.3\times10^{11}$~\lsun\ obtained by \cite{wilson14}. The luminosity of the east nucleus is consistent with the $\sim$~1/3 of the total luminosity allocated by \cite{scoville17} to this source.

Most of the IR luminosity from Arp~220 arises from the diffuse circumnuclear region. The equilibrium temperature from the diffuse component is about 40~K, about twice as high as that of the dust in the local diffuse ISM.  The average luminosity per unit mass in the diffuse medium of Arp~220 is $\sim 90$~\lsun/\msun, about 15 times higher than the average value of 6~\lsun/\msun\ in the molecular clouds in the Milky Way \citep{hauser84}. 
\subsection{Star formation rates}
   
These luminosities can be converted into star formation rates (SFR), provided that they are powered by star formation activity. The average luminosity expended in the formation of 1~\msun\ of stars per year depends on the stellar initial mass function (IMF) and the age of the starburst, and is equal to $\sim 1.1\times10^{10}$~\lsun\ for a Kroupa IMF and a starburst age of 100~Myr \citep{dwek11b}. If the radiated IR emission is powered by star formation then the inferred SFR for the east and west nuclear components and the diffuse medium are  32, 53, and (46--62)~\myr, respectively. These values are consistent with the value of $\sim 100$~\myr\ for the whole galaxy \citep{scoville03}. Given these SFRs, the gas reservoir in the nuclei will be consumed in about 20--100~Myr, depending on the adopted gas mass.

An important quantity characterizing the nature of a star forming region is its $L/M$ ratio. A star forming region is fueled by the collapse or infall of ambient gas. However, in an extremely dusty environment the accretion of gas can be halted by the stellar radiation pressure on the infalling dust. The parameter $\beta$ is defined as the ratio of the radiation force to the gravitational force on a grain, where $\beta=1$ corresponds to the Eddington limited luminosity for accretion
\begin{eqnarray}
\label{beta}
\beta \equiv {F_{pr}\over F_{grav}} & = & \left[{L\over 4\pi R^2\ c}\, \pi a^2 Q_{pr}\right] \times \left[{G\, m\, M \over R^2}\right]^{-1} \nonumber \\
& = & \left({L\over M}\right)\, {\kappa_{pr}\, Z_{dH}\over 4\pi\, G\, c} \qquad , 
\end{eqnarray}  
where $L$ and $M$ are the luminosity and mass of the star forming region, $G$ is the gravitational constant, $m$ the mass of an infalling parcel of gas,  
\begin{equation}
\label{Qpr}
Q_{pr} = Q_{abs}+Q_{sca}\, [1-\left<cos(\theta)\right>]
\end{equation}
is the radiation pressure efficiency of the dust, where $\left<cos(\theta)\right>$ is its scattering asymmetry parameter, and $\kappa_{pr} = \pi a^2\, Q_{pr}/m_{gr}$ is the radiation pressure mass absorption coefficient. The value of $\left<cos(\theta)\right>$ is equal to -1, 0, +1 for completely backscattering, isotropically scattering, and completely forward scattering grains. 

Infall is halted if $\beta > 1$, or  
\begin{equation}
\label{lm}
{L\over M} > \left({L\over M}\right)_{Edd}  \equiv   {4 \pi G\, c\over \kappa_{pr}\, Z_{dH}} \qquad .
\end{equation} 
The effective value of $\kappa_{pr}$  depends on the radiation field acting on the dust. In optically thin layers most of the radiation pressure will be provided by direct starlight. At visible wavelengths $\kappa_{pr} \approx$ 230~cm$^2$~g$^{-1}$. However, in optically-thick regions the radiation pressure will be provided by dust-reprocessed starlight at near- to mid-IR wavelengths \citep{crocker18}. The value of $\kappa_{pr}$ ranges from 5 to 10~cm$^2$~g$^{-1}$ in the 5 to 20~\mic\ wavelengths region. All above values for $\kappa_{pr}$ were calculated for a dust-to-H mass ratio of 0.007 and a 4:3 mass ratio of 0.1~\mic\ radius silicate to carbon grains. \cite{scoville17} adopted a value of $\kappa \approx 10$~cm$^2$~g$^{-1}$ at near-IR wavelengths. The limiting $L/M$ ratio at visual (0.55~\mic), 10, and 20~\mic\ spans a large range of values given (in solar units) by  
\begin{eqnarray}
\label{lm2}
\left({L\over M}\right)_{Edd} & \approx & \ 55,\ 1300,\ {\rm and}\ 2800 \nonumber \\ 
& & {\rm at}\ 0.55,\ 10,\ {\rm and}\ 20~\mu m
\end{eqnarray}
Deriving a more accurate value for $({L/M})_{Edd}$ is much more complicated, requiring a detailed radiative transfer model to derive the radiation pressure on the dust throughout the gas.  

The observed $L/M$ ratio is somewhat uncertain primarily because of the mass in the disk. The average luminosity of a nucleus is $\sim 4.5\times10^{11}$~\lsun. If the IR luminosity is completely generated by the accretion on a central BH, then the central mass is about $10^7$~\msun (see Section~5.7), giving a $L/M$ value of $4.5\times10^4$, in solar units. If all the IR luminosity is powered by star formation activity then the disk mass is given by the dynamical stellar mass of the system. The estimate for this mass ranges from $8\times 10^8$~\msun\ for a point-like stellar distribution to $1.5\times 10^9$~\msun\ for  an extended disk-like distribution. Following \cite{scoville17}, we adopt a value of $10^9$~\msun\ for the mass. The resulting $L/M$ value is then 450, compared to their value of 500. 

The observed $L/M$ ratio is smaller than that for an optically-thick star forming region. Radiation pressure is therefore not the dominant force driving the observed gas outflows from the nuclei, which must therefore be mostly driven by stellar winds and SN shock waves \citep{paggi17}.

\subsection{\cii as a SFR tracer in optically thick systems}
The detection of the \cii 158\mic\ fine structure line in the eastern nucleus of Arp~220 provides an opportunity to examine the applicability of this line as a tracer of the SFR in this region. The \cii\ line correlates with other \neii 12.8\mic\ and \neiii 15.6\mic\ cooling lines that arise from H~II regions, and with the total IR luminosity in normal SF galaxies. The \cii\ luminosity has therefore been suggested as tracer of the SFR in galaxies \citep[][and references therein]{kennicutt12}. However, observations have shown a trend of increasing \cii\ deficit with IR luminosity, especially in ULIRGs \citep{malhotra01, luhman03}. The deficiency has been attributed to a high FUV flux-to-gas density ratio, expressed as $G_0/n$ \citep{hollenbach97, malhotra01}; a high ionization parameter $U$, defined as the ratio of the hydrogen-ionizing photon density-to-hydrogen density \citep{gracia-carpio11,fischer14}; or the effect of dust obscuration \citep{farrah13, diaz-santos13}. These interpretations were challenged by \cite{sargsyan14} who claimed that the deficit does not reflect the decrease of the \cii-to-starburst generated luminosity, but rather the increased contribution of an AGN to the total IR luminosity. Their conclusion  is based on the fact that the more luminous IR sources with $L_{IR} \gtrsim 10^{12}$~\lsun\ have also weak PAH and \neii\ and \neiii\ line emission, indicating that they are harboring an AGN. Since the near- and mid-IR emission arises primarily from hot AGN-heated dust, they suggested the $L$(\cii) to the $\nu L_{\nu}$(158~\mic) continuum ratio, which arises from cooler dust heated by active star formation, as an alternate tracer of the SFR in galaxies with mixed AGN and starbursts (SB) as power sources. Based on various calibrators, \cite{sargsyan14} derived a $L$(\cii)-to SFR conversion factor given by $\log$(SFR) = $\log(L$\cii)$ - 7.0 \pm 0.2$, where the SFR and $L$\cii) are in units of \myr\ and \lsun, respectively.
     
None of their arguments are applicable to systems that are optically thick at far-IR wavelengths. All PAH and fine structure lines are extinguished by dust, with opacities that increase with decreasing wavelengths. Furthermore, at the optical depths encountered in the nuclei of Arp~220, AGN as well as SB will contribute to the IR emission across all wavelengths. 
The total luminosity in the \cii\ line from the eastern nucleus is about $8.2\times 10^7$~\lsun \citep{samsonyan16}\footnote{\url{https://cassis.sirtf.com/herschel}}, giving a SFR of $8.2^{+4.7}_{-3.0}$~\myr, compared to the value of 32~\myr\ derived here from the total IR luminosity. This large discrepancy is consistent with the fact that the \cii\ line is heavily obscured. Figure \ref{sedfits} shows that the 158~\mic\ radial optical depth for the east nucleus is $\sim$~25, and the line can only be probed down to a radial distance of about 45~pc. The \cii\ line intensity can therefore not be used as an indicator for the star formation activity taking place in the center of the nucleus.  

\subsection{Surface densities}
\begin{figure}[t]
\includegraphics[width=3.0in]{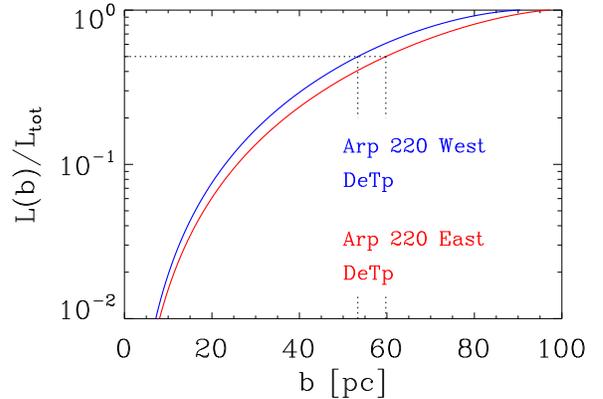}
\caption{\label{Lconvol}  The fractional luminosity emerging from the projected area within the impact radius $b$. Half the luminosity from Arp~220~East and West originates within a radius of 59 and 53~pc, respectively. }
\end{figure}

Important quantities that shed light on the nature of galaxies are the surface densities of the luminosities, the SFR, and the gas mass. We define the effective area for calculating these quantities as the area from which half the sources' luminosity is emitted.
Figure~\ref{Lconvol} shows the cumulative fraction of the observed nuclear luminosity as a function of the impact parameter $b$. The figure shows that half of the sources' luminosities originate within  radii $R_{1/2}$ of 53, and 59~pc for the west and east nucleus, respectively.

The luminosity per unit area, $\Sigma_L$, calculated for the area $\pi\, R_{1/2}^2$,  is $3.3\times 10^{13}$ and $3.2\times 10^{13}$~\lsun~kpc$^{-2}$, for the west and east nuclei, respectively.
The corresponding SFR per unit area, $\Sigma_{SFR}=\Sigma_L/1.1\times 10^{10}$, and the gas mass surface density, are listed in Table~1.  

\cite{thompson05} studied the conditions that will lead to a stable star forming disk in which radiation pressure on the dust supports the disk against gravity.  In optically thick star forming regions the relation between the luminosity and the SFR breaks down when the luminosity exceeds the Eddington luminosity for dust. The Schmidt-Kennicutt relation is then modified by the disk opacity $\tau$ so that  $\Sigma_{SFR} \propto \Sigma_{gas}/\tau$. \citep{thompson05}. The resulting characteristic surface densities required to stabilize the disk are   $\Sigma_L \approx 10^{13}$~\lsun~kpc$^{-2}$, $\Sigma_{SFR} \approx 10^3$~\msun~yr$^{-1}$~kpc$^{-2}$, and a dust temperatures of $\sim 90$~K \citep{thompson05}. These  values are close to those of the two nuclei, suggesting that radiation pressure plays an important role in limiting the fueling of the star formation activity in the nuclei.  

\subsection{Metal and dust enrichment by supernovae}
The potentially large dust-to-H mass ratios in the nuclei raises the possibility that this dust and associated metal enrichment, could have been caused by supernova activity.  From the observed luminosities and inferred SFR of the various components, we can derive the average supernova rate (SNR). The mass of stars formed per SN event is 90~\msun\ for a Kroupa stellar initial mass function (IMF) \citep{dwek11b}. The average supernova rate (SNR) in the nuclei is therefore $\sim 0.5$~yr$^{-1}$. Using  the nucleosynthesis yields of \citep{woosley07}, the IMF-averaged mass of metals (all elements heavier than helium) injected by SNe is about 4~\msun, of which 0.6~\msun\ are locked up in dust. If all the dust survives its injection into the ambient medium, then the average  metal and dust enrichment rates are about 2 and 0.3~\msun~yr$^{-1}$, respectively. The total mass of heavy elements and dust injected into the ISM in a period of $10^7$~yr is then about $2\times 10^7$~\msun\ and $3\times 10^6$~\msun, respectively. SN ejecta can therefore play an important role in the enrichment of the ambient gas. However, these masses represent upper limits on the metal and dust enrichments, since a significant mass of SN produced metals and dust can be ejected from the nuclei by the hot outflows, and a significant fraction of the dust mass can be destroyed by shocks in the violent starburst environment. 

\subsection{The obscured black hole}
So far we have assumed that all the luminosity is powered by star formation in the center of the two nuclei. If the luminosity is powered by an accreting black hole (BH), then the paucity of  observed hard X-rays from the nuclei suggests that it must be deeply enshrouded by a large column density of gas \citep[][and references therein]{iwasawa05}. The hydrogen column density, as inferred from the infrared observations, is given by
\begin{equation}
\label{tauH}
N_{H} = {\tau_R(\lambda)\over Z_{dH}\, m_H\, \kappa_d(\lambda)}
\end{equation}  
At 100~\mic\ the dust mass absorption coefficient is equal to 41~cm$^2$~g$^{-1}$, and almost identical in value for both the  compositional mixes in the nuclei (Fig. \ref{kappa}, right panel). The radial optical depth of the two nuclei at this wavelength is 37.5 and 62.7 for the west and east nuclei, respectively. For a dust-to-H mass ratio of 0.007 the value of $N_H$ is equal to $7.9\times 10^{25}$ and $1.3\times 10^{26}$~cm$^{-2}$, for the west and east nuclei, respectively. These column densities are comparable to those derived by \cite{scoville17}. They are well above the minimum value of $\sim 10^{25}$~cm$^{-2}$ required to account for the paucity of hard X-ray photons. 

 The mass of a BH accreting at the Eddington luminosity \citep{rybicki86} is given by $M_{BH} = 3\times 10^6\, L_{11}$~\msun, where $L_{11}$ is the IR luminosity in units of $10^{11}$~\lsun. The accretion rate is given by $\dot{M} = 2.3\times10^{-9}\, \epsilon^{-1}\, M_{BH}$~\msun~yr$^{-1}$, where the BH mass is in solar unit, and $\epsilon$ is the radiative efficiency of the accretion process. For an average IR luminosity of $\sim 4.5\times 10^{11}$~\lsun, the BH mass is about $1.3\times 10^7$~\msun, with a current accretion rate of 0.11~\msun~yr$^{-1}$ for $\epsilon=0.3$. The accretion lifetime is about 130~Myr, depleting half the reservoir of gas in 90~Myr.  
   
\subsection{The nature of the warm dust component}
The need for two distinct components to fit the entire IR to submillimeter SED suggests  that a single continuous distribution of dust density, whether uniform, exponential or power law cannot fit both the long and short wavelength emission, especially the 10~\mic\ absorption features observed in the two nuclear regions of the galaxy. However, the need for two distinct components does not imply the presence of two distinct heating sources, but rather illustrates the complex morphology of the nuclei. For example, the 10~\mic\ absorption feature could represent a line of sight that probes a deep layer of the source, where the dust temperature is around 320~K, which has an intervening column density of cold dust that  gives rise to the 10~\mic\ absorption feature.
 
\subsection{Dust compositions}
An interesting result was the need for different combinations of the dust compositions to fit the long wavelength emission from the two nuclei. The spectrum of Arp~220~West has a shallower $\lambda^{-2.4}$ wavelength dependence, compared to the $\lambda^{-3.1}$ dependence of Arp~220~East. In principle one could fit both nuclei with the same combination of SIL+GRF grains and add a cold dust component to Arp~220~West. However, such dust component will require $ > 10^9$~\msun\ of dust radiating at temperatures around $\sim 3$~K to fit the flatter spectrum of the west nucleus. The difference in the nature of the carbonaceous dust component in the two nuclei is unclear, with different nucleation environments or cosmic ray exposure histories of the two galaxies as possible explanations.

\section{Summary}
Our analysis has provided new physically derived estimates of the dust masses, luminosities, and dust temperatures in the nuclear and diffuse circumnuclear components of Arp~220. 
We find that the nuclear regions are deeply shrouded in dust with radial optical depths of unity at 1890 and 775~\mic\ for the west and east nucleus, respectively.
  
It has revealed a significant discrepancy between the dust inferred gas mass and the CO inferred molecular gas mass, suggesting significantly larger dust-to-H mass ratios in the nuclei compared to that in the local ISM, or the presence of atomic gas that is undetected by the CO observations. 

Supernovae can make a significant contribution to the observed metallicity and dust content of the nuclei. However this contribution may be limited by stellar winds and dust destruction processes.

The central sources powering the IR emission from the two nuclei are embedded in a region with a large H-column density of $\sim 10^{26}$~cm$^{-2}$, with an exponential density profile. However, the nature of the central sources cannot be determined from our analysis. An average SFR of $\sim 50$~\myr\ can generate the observed IR luminosity from each nucleus and consume its ambient gas reservoir  in about 10~Myr. 

The observed \cii\ emission from Arp~220~East is heavily obscured and only arises from the outer region of the nucleus. It can therefore not be used as a calibrator of any star formation activity in its center. 

The average nuclear $L/M$ ratio is smaller than the Eddington limit for a dusty star forming region suggesting that the observed outflows are not driven by radiation pressure.
However, the observed luminosity and SFR surface densities are close to the theoretical values required to counteract the effects of gravity by radiation pressure on the dust.    
 
Alternatively, deeply obscured accreting black holes with average masses of $\sim 1\times 10^7$~\msun\ can give rise to the observed luminosity from each nucleus. With a radiative accretion efficiency of 0.3 its accretion lifetime is about 130~Myr, depleting half the reservoir of gas in each nucleus in about 90~Myr. 

The short lifetime of the IR activity in Arp~220 is an important factor in determining the number of such superluminous system in the universe.    

\begin{deluxetable*}{llll}
\decimals
\tablewidth{0pt}
\tablecaption{Summary of the best-fit parameters for the two nuclei and the diffuse emission components of Arp~220}
\tablehead{
\colhead{ } &
\colhead{East (DeTp)} &
\colhead{West (DeTp)} &
\colhead{Diffuse}\\
\colhead{ } &
\colhead{(SIL+GRF)} &
\colhead{(SIL+ACAR)} &
\colhead{(SIL+ACAR)}
 }
 \startdata
    cold component & & & \\
   \hline     
  $M_d$ [\msun]    & $3.4\times10^{7}$     &    $1.8\times10^{7}$     &  $\sim 8.0\times10^{7}$   \\ 
  $T_d$ [K]        &  171                  &     199                  &    $\sim$ 40       \\
   $a$ [pc]                           &  13.9                 &    12.9                  & \nodata      \\
   $L_d$ [\lsun]                      & $3.5\times10^{11}$    &     $5.8\times10^{11}$   &    $\sim 9.5\times10^{11}$  \\
   $R_{1/2}$ [pc]                    &  59                &    53                  & \nodata      \\
   $\tau_R$(2600~\mic)              &  0.09                 & 0.72          & \nodata \\
    $N_H$ [cm$^{-2}$]                    &  $1.3\times 10^{26}$    &   $7.7\times 10^{25}$                  & \nodata      \\
    SFR [\msun~yr$^{-1}$]             &  32                   &    53                     &    46\, --\, 62  \\
    SNR [yr$^{-1}$]                   &  0.36                   &    0.59                     &    0.5\, --\, 0.7  \\
   $\lambda(\tau=1)$ [\mic]                &   775           &     1890           &   \nodata   \\
    $M_{gas}$ [\msun]                 & $4.9\times10^{9}$           &    $2.9\times10^{9}$        &  $\sim 1.1\times10^{10}$   \\ 
    $L_d/M_{gas}$ [\lsun/\msun]         &  71                    &   242                  & \nodata     \\
   $\Sigma_{gas}$ [\msun~pc$^{-2}$]  &   $4.5\times10^{5}$      &    $2.9\times10^{5}$     &   \nodata   \\
    $M_{H_2}$ [\msun]                 & $(0.4-0.8)\times10^{9}$     &    $(0.8-2.4)\times10^{9}$  &  \nodata   \\
    $\Sigma_{H_2}$ [\msun~pc$^{-2}$]  &   $5.5\times10^{4}$      &    $1.8\times10^{5}$     &   \nodata   \\
   $L_d/M_{H_2}$ [\lsun/\msun]         &  583                    &   320                  & \nodata     \\
   $\Sigma_L$ [\lsun~kpc$^{-2}$]     &   $3.2\times10^{13}$       &    $3.3\times10^{13}$     &   \nodata   \\
   $\Sigma_{SFR}$ [\msun~yr$^{-1}$~kpc$^{-2}$]  &   $2900$      &    $3000$             &   \nodata   \\
    \hline 
   warm component & & & \\
   \hline  
   $R_{bb}$ [pc]                      &   1.7                 &     1.7                  &   \nodata   \\
   $T_{bb}$ [K]                       &   307                 &     338                  &   \nodata   \\
   $\Sigma_d$\, [g\ cm$^{-2}]$            &   $2.3\times10^{-3}$  &     $1.5\times10^{-3}$   &   \nodata   \\
   $L_{bb}$ [\lsun]                   &   $4.6\times10^{10}$  &     $6.8\times10^{10}$   &   \nodata   \\
   $L_d$ [\lsun]                      &   $1.1\times10^{10}$  &     $2.3\times10^{10}$   &   \nodata   \\
   $\tau$(10 \mic)                &   4.2                 &    3.1                  &   \nodata   \\
   $\tau$(20 \mic)                &   1.45                 &      1.00                &   \nodata   \\
  \enddata
  \tablecomments{ Dust temperature for the West nucleus corresponds to the central value in the exponential profile. 
  $R_{1/2}$ is the radius of the area giving rise to half of the observed luminosity (Fig. \ref{Lconvol}); $\tau_R$(2600~\mic) is the radial optical depth at 2600~\mic\ (eq.~(\ref{taur})); $N_H$ is the radial H-column density derived from eq.~(\ref{tauH}); 
Star formation rates (SFR) were derived from the luminosities for a Kroupa IMF $L$/SFR ratio of $1.1\times 10^{10}$ for the nuclei, and a ratio of ~$(1.53-2.08)\times 10^{10}$~\lsun/\msun~yr$^{-1}$ for the diffuse component. Supernova rates (SNR) were derived from the SFR for an IMF-averaged star or 90~\msun\ per SN event. Surface densities are calculated for the area of $\pi R_{1/2}^2$,  $M_{gas}$ is the dust inferred gas mass for a dust-to-gas mass ratio of 0.007. $M_{H_2}$ is the CO inferred mass of molecular gas for an $X$ factor of 0.8 \citep{wheeler20}.}
  \label{table2}
\end{deluxetable*}

\acknowledgements
The research for this project was supported by NASA 16-ATP16-0004. We thank the referee for pointing out the detection of the \cii\ emission in Arp~220~East, and for suggesting the discussion on the applicability of this line as a SFR tracer in heavily obscured systems. We also acknowledge helpful discussions with Jackie Fischer, Jason Glenn, and Nick Scoville.  


\clearpage
\bibliography{/Users/edwek/OneDrive-NASA/science/00-Bib_Desk/Astro_BIB.bib}

 \end{document}